\documentclass[twocolumn,showpacs,preprintnumbers,amsmath,amssymb]{revtex4}
\usepackage{graphicx}

\def\la{\mathrel{\hbox{\rlap{\hbox{\lower4pt\hbox{$\sim$}}}\hbox{$<$}}}}
\def\ga{\mathrel{\hbox{\rlap{\hbox{\lower4pt\hbox{$\sim$}}}\hbox{$>$}}}}

\begin{document}

\title{Hot plasma in clusters of galaxies, the largest objects in the
universe}

\author{Craig L. Sarazin}
\affiliation{Department of Astronomy, University of Virginia,
530 McCormick Road, Charlottesville, VA 22903-0818}

\date{\today}

\begin{abstract}
Clusters of galaxies are the largest organized structures in the Universe.
They are important cosmological probes, since they are large enough to
contain a fair sample of the materials in the Universe, but small enough to
have achieved dynamical equilibrium.
Clusters were first discovered as concentrations of hundreds of bright
galaxies in a region about 3 megaparsecs (10 million light years) across.
However, the dominant observed form of matter in clusters is hot, diffuse
intergalactic gas.
This intracluster plasma has typical temperatures of $T \sim 7 \times 10^7$ K,
and typical electron densities of $n_e \sim 10^{-3}$ cm$^{-3}$.
This intracluster plasma mainly emits X-rays, and typical
cluster X-ray luminosities are $L_X \sim 10^{43} - 10^{45}$ erg/s.
The basic properties of and physical processes in the intracluster plasma will
be reviewed.
Important observational constraints on plasma processes in these systems will
be discussed.
Recent X-ray observations of clusters of galaxies with the orbiting
Chandra X-ray Observatory will be highlighted.
\end{abstract}

\pacs{95.30.Qd,98.65.Cw,98.65.Hb}

\maketitle

\section{Introduction} \label{sec:intro}

\begin{figure*}
\vskip3.40truein
\includegraphics{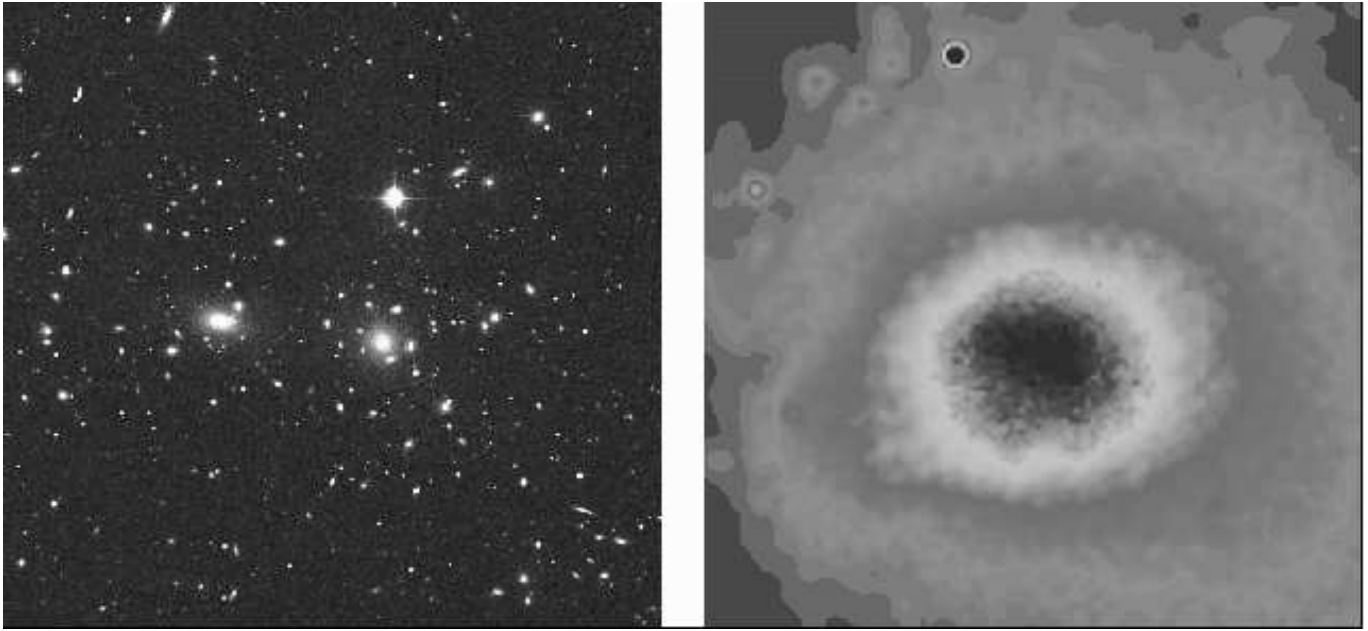}
\caption{\label{fig:fig1} The left panel shows an optical image of the
central region of the nearby Coma cluster of galaxies, showing the
galaxies in the cluster.  The right panel shows the same region
imaged in X-ray emission with the R\"ontgen Satellite (ROSAT).
The X-ray emission comes from hot plasma at about $10^8$ K which fills the
volume of the cluster.}
\end{figure*}

Clusters of galaxies are the largest relaxed structures in the Universe.
They were initially discovered in optical observations, where clusters
appear as concentrations containing $\sim 10^2$ bright galaxies and
$\sim 10^3$ faint galaxies in a region which is typically $\sim 2$
Mpc in radius (1 megaparsec $= 3.09 \times 10^{24}$ cm) \cite{aco89}.
For example,
Figure~(\ref{fig:fig1}, left panel) shows the optical image of the
central region of the nearby Coma cluster showing many galaxies.
In their central regions, clusters are about $10^3$ times denser than
the average of material in the Universe.
Clusters of galaxies are very important cosmological probes \cite{nab00}.
Essentially, they are the only objects in the Universe are both
small enough to have achieved dynamical equilibrium during the age of the
Universe, and large enough to contain a fair sample of the material
in the Universe (e.g., the average ratio of baryonic to dark matter).

Although they were first observed as collections of galaxies,
the dominant form of matter which has been observed in clusters of
galaxies is hot diffuse plasma \cite{sar86}.
This intracluster medium (ICM) has typical temperatures of
$\sim 7 \times 10^7$ K and typical electron number densities of 
$n_e \sim 10^{-3}$ cm$^{-3}$.
At these temperatures,
the dominant form of radiation from a plasma
is X-ray emission, mainly from thermal bremsstrahlung but also
from collisionally excited line emission.
As a result, clusters of galaxies are generally very luminous X-ray
emitters,
with luminosities of $L_X \sim 10^{43} - 10^{45}$ ergs s$^{-1}$.
Clusters are second only to quasars as the most luminous X-ray sources
in the Universe.
For example, Figure~(\ref{fig:fig1}, right panel) shows the X-ray image of the
same central region of the Coma cluster as the left panel.
Although the ICM is diffuse, it fills all of the volume between and within
the galaxies in clusters, and as a result its mass is
large.
The total mass of hot plasma in a cluster is typically
$M_{gas} \sim 10^{14} \, M_\odot$, where $M_\odot = 1.99 \times 10^{33}$ g
is the mass of the Sun.
In large clusters, the total mass of hot gas
exceeds the mass of all the stars and galaxies by a factor of $\sim$5.
Hot intracluster plasma is the dominant form of baryonic
matter in clusters.
In general, we now believe that most of the baryonic matter in the low
redshift Universe is in the form of hot intergalactic plasma.

\section{Physical Properties of the Intracluster Plasma} \label{sec:ICM}

The mean free paths of electrons and ions in a plasma without a magnetic
field are determined by Coulomb collisions.
The mean free path of electrons (which is nearly the same as that protons)
is \cite{spi56}
\begin{equation}\label{eq:mfp}
\lambda_e = \frac{ 3^{3/2} (k T )^2 }{4 \pi^{1/2} n_e e^4 \ln \Lambda }
\approx 23 
\left( \frac{T}{10^8 \, {\rm K}} \right)^2
\left( \frac{n_e}{10^{-3} \, {\rm cm}^{-3}} \right)^{-1} \, {\rm kpc} ,
\end{equation}
where  $n_e$ is the electron number density, the Coulomb logarithm
$\ln \Lambda \approx 38$, and 1 kpc $= 3.09 \times 10^{21}$ cm.
This is about 1\% of the radius of a cluster, which suggests that the
intracluster plasma can be treated as a fluid.
The gyroradii in the intracluster magnetic field are much smaller than
this (Sec.~\ref{sec:magnetic}).

The timescale for Coulomb collisions between electrons to bring them
into kinetic equilibrium (an isotropic Maxwellian velocity distribution)
is about \cite{spi56}
\begin{equation}\label{eq:tee}
t_{\rm eq} (e,e) \approx 3.3 \times 10^5
\left( \frac{T}{10^8 \, {\rm K}} \right)^{3/2}
\left( \frac{n_e}{10^{-3} \, {\rm cm}^{-3}} \right)^{-1} \, {\rm yr} \, .
\end{equation}
The time scale for protons to equilibrate among themselves is
$t_{\rm eq} (p,p) \approx ( m_p / m_e )^{1/2}$ $t_{\rm eq} (e,e)$
or roughly 43 times longer than the value in Equation~(\ref{eq:tee}).
Similarly, time scale for the electrons and ions to reach equipartition
(equal temperatures) is
$t_{\rm eq} (p,e) \approx ( m_p / m_e ) t_{\rm eq} (e,e)$,
or roughly 1870 times the value in Equation~(\ref{eq:tee}).
All of these are shorter than the typical ages of clusters of $\ga 10^9$
yr.
Thus,  the intracluster plasma can generally be characterized by a Maxwellian
distribution at a kinetic temperature $T$.
Because the radiation field is much more diffuse than a blackbody at
the kinetic temperature, the gas if far from thermodynamic equilibrium
in terms of the populations of excited levels of ions and the ionization
state.
However, the gas is in ``coronal equilibrium'';
most bound electrons are in their ground levels, they are in excitation
equilibrium, and the gas is in ionization equilibrium.

The sound crossing time for a cluster is
\begin{equation}\label{eq:tsound}
t_s \equiv \frac{D}{c_s} \approx 6.6 \times 10^8
\left( \frac{T}{10^8 \, {\rm K}} \right)^{-1/2}
\left( \frac{D}{{\rm Mpc}} \right) \, {\rm yr} \, ,
\end{equation}
Here, $D$ is the diameter of the cluster, and $c_s$ is the sound speed.
This is somewhat smaller than the likely ages of clusters, so unless
they are being disturbed (Sec.~\ref{sec:mergers}), the gas should
be nearly in hydrostatic equilibrium.
For a spherical cluster in hydrostatic equilibrium,
the gas distribution is given by
\begin{equation}\label{eq:hystatic}
\frac{1}{\rho_{\rm gas}} \frac{dP_{\rm gas}}{dr} =
- \frac{G M (r)}{r^2} \, ,
\end{equation}
where $M (r)$ is the total cluster mass within a radius $r$.
The gas pressure in the ICM is given by the ideal gas law,
$ P_{\rm gas} = \rho_{\rm gas} k T / ( \mu m_H )$,
where $ \rho_{\rm gas}$ is the mass density in the gas, and
$ \mu m_H $ gives the mean mass per particle.
One often assumes that magnetic forces, pressure from relativistic particles,
and other forces are relatively weak in clusters, although it is not
certain that this is correct
(Sec.~\ref{sec:magnetic}).
Equation~(\ref{eq:hystatic}) has been used to determine the total masses
or total density profiles of clusters by solving for $M( r )$.
The temperature of the ICM can be determined from observations of
the X-ray spectrum of the gas, while the X-ray surface brightness
can easily be de-projected to give the gas density.
Such measurements indicate that the total masses of large clusters
are about $10^{15} \, M_\odot$, which considerably exceed the total
mass of all of the intracluster gas and of all the galaxies combined.
As such, clusters provide some of the strongest evidence for the
domination of (probably nonbaryonic) dark matter on large scales in
the Universe.
In a typical large cluster, $\sim$16\% of the mass is in hot ICM,
$\sim$3\% of the mass is in stars and galaxies, and $\sim$81\% of
the mass appears to be dark matter.

As is true of most materials in the Universe, the ICM consists primarily
of ionized hydrogen ($\approx 71$\% of mass) and
helium ($\approx 28$\% of mass).
However, the ICM does contain a significant amount of the common heavier
elements (O, Fe, etc.; $\approx 1$\% of mass).
This fraction is only about a factor of 2--3 times smaller than the
fraction in the Sun.
Many of these heavy elements are detected through X-ray lines observed
in the spectra of clusters of galaxies; these lines occur because the
heavier elements are not quite completely ionized, even at the high
temperatures in clusters.
Because most of the baryonic matter in clusters is in the ICM,
it turns out that most of the heavy elements are actually located there
as well.
Hydrogen and helium are formed in the Big Bang, but the only source of
the common heavier elements is fusion reactions in the centers of
stars.
At present, the only significant populations of stars are located in
galaxies.
The dispersal of heavy element into the diffuse intracluster gas required
that stars in galaxies be disrupted by supernova explosions, and that the
enriched gas escape the gravity of the galaxy in which the stars are
located.
Detailed models for the chemical evolution of the ICM suggest that
about 25\% of the gas originated in stars in galaxies, and that the
remaining 75\% came from primordial intergalactic gas.
Because there is presently about five times as much mass in the ICM as
in stars and galaxies, this requires that the galaxies located in
clusters lost significant amounts of their baryonic content.
At present, the galactic population in clusters of galaxies consists mainly
of elliptical (E) and lenticular (S0) galaxies, which have only low mass
stars which do not produce a high rate of supernovae.
To explain the large amounts of heavy elements in the ICM, these
galaxies must have had much higher rates of star formation and supernovae
at earlier times, possibly associated with the formation of the galaxies.

Initially, the very high temperature ($\sim 10^8$ K) of the ICM might seem
surprising.
However, this is one feature of clusters which is easily understood.
Clusters of galaxies contain enormous masses of material, and have very
deep gravitational potential wells.
Almost any natural process which introduces gas into clusters will
cause it to move very rapidly and be shock-heated to roughly the
observed temperature.
For example, if the ICM fell into clusters, it would be accelerated to
roughly the escape speed from clusters, which is $\sim 2000$ km/s.
A portion of the ICM may have been ejected from galaxies;
galaxies in clusters move on random orbits with velocities of
$\sim 1000$ km/s.
In either case, infalling or ejected gas would encounter other gas
moving at
similar velocities, and would be undergo strong hydrodynamical shocks.
Shocks at speeds of $\sim 1000$ km/s heat gas to temperatures of
$\sim 10^8$ K.

\section{Magnetic Fields and Relativistic Particles} \label{sec:magnetic}

Although the ICM is dominated in mass and energetically by thermal
plasma, it does contain magnetic fields and populations of relativistic,
nonthermal particles as well.
The most direct way to measure the magnetic field in clusters of galaxies
is through the Faraday rotation of the polarization of background
or embedded radio sources \cite{ckb01}.
Most strong extragalactic radio sources emit synchrotron radiation which
is strongly linearly polarized.
When this radiation passes through a magnetized plasma,
the plane of polarization is rotated through an angle
\begin{equation}\label{eq:faraday}
\phi = (RM) \lambda^2 \, ,
\end{equation}
where $\lambda$ is the wavelength of the radiation.
The rotation measure, $RM$, is given by
\begin{equation}\label{eq:rm}
RM =  \frac{e^3}{2 \pi m_e^2 c^4 } \int n_e B_\| dl
\, ,
\end{equation}
where $l$ is the path length through the medium, and $B_\|$ is the
component
of the magnetic field parallel to the direction of propagation of the
radiation.
The rotation measures seen through the central regions of clusters
are $RM \sim 100$ rad m$^{-2}$.
Unfortunately, Equation~(\ref{eq:rm}) shows that the rotation measure only
determines an integral of $B_\|$ along the line of sight. and this
integral is strongly affected by the poorly known topology of the
magnetic field.
Field reversals along the line of sight will greatly reduce $RM$.
If one assumes that the coherence length of the field is about 10 kpc,
the rotation measure observations suggest that the typical magnetic field
strength is $B \sim 5 \, \mu$G \cite{ckb01}.
This would imply that the fields are still significantly subthermal; the ratio
of magnetic to gas pressures is only $( P_B / P_{\rm gas} ) \sim 0.05$.
In the central regions of clusters with cooling cores
(Sec.~\ref{sec:centers}),
much larger rotations measures are observed ($RM \sim 10^4$ rad m$^{-2}$),
which suggest that the magnetic fields in these regions may approach
equipartition with the gas pressure.

\begin{figure}
\includegraphics{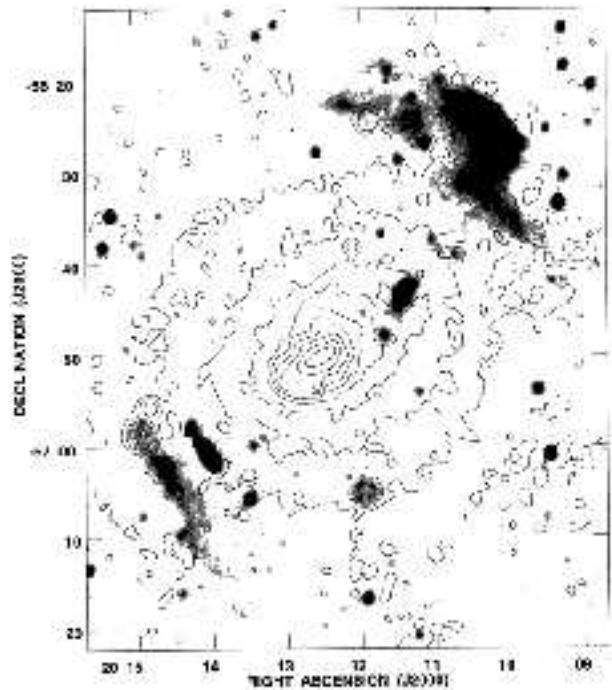}
\caption{
\label{fig:fig2}
The grey-scale is the radio image of the
cluster Abell~3667 \cite{rwh+97}, 
while the contours are an X-ray image.
The radio image shows two extended arcs, which are cluster radio relics,
located to the southeast (lower left) and to the northwest (upper right) of
the cluster center.
These radio relics are believed to be associated with cluster merger
shocks.
(\copyright 1997, The Royal Astronomical Society.)
}
\end{figure}

In addition to the thermal plasma in the intracluster medium,
significant population of relativistic electrons are observed in
some clusters.
The observations are crudely consistent with a power-law distribution
for the relativistic particles;
that is, $N_e ( E ) \propto E^{-p} $, where $N_e ( E ) \, dE$ gives the
number of electrons with energies in the range $E \rightarrow E + dE$.
Similar distributions are seen in other astrophysical plasmas including
the Galactic cosmic rays.
However, it is likely that the particle distributions are more
complex, and also vary spatially.
Relativistic electrons interact with the intracluster magnetic field
to produce synchrotron radio emission;
the electrons which produce the emission typically have energies
$\sim 10$ GeV.
Diffuse emission, not associated with any individual galaxy,
is seen in $\sim$40 clusters of galaxies;
when the emission is centrally located, the sources are called
``cluster radio halos,'' while ``cluster radio relics'' are peripherally
located \cite{gf02}.
For example, Figure~(\ref{fig:fig2}) shows two radio relic sources in the
cluster Abell~3667 \cite{rwh+97}.

Relativistic electrons in clusters can also produce observable emission
through the inverse Compton scattering of low energy photons;
the main source of these photons in clusters of galaxies is the
Cosmic Microwave Background.
Typically, inverse Compton scattering produces emission which is observable
either in the extreme-uv/soft X-ray band (near 0.1 keV) or in the the hard
X-ray band (20--100 keV).
Recently, inverse Compton hard X-ray emission has been detected from
several clusters \cite{ff+99},
Extreme-uv/soft X-ray emission, which might also be from relativistic
electrons, may have also been detected \cite{sl98}.
Detecting both inverse Compton and synchrotron emission from the same
population of relativistic electrons is useful, because the synchrotron
emission depends on the product of the energy density in relativistic
electrons and that in the magnetic field, while inverse Compton emission
depends on the product of the energy density in relativistic
electrons and that in the Cosmic Microwave Background (which is very
well determined).
In principle, the combination of these two measurements allows both the total
energy in relativistic electrons and the magnetic field strength to be
determined.
However, one needs to assume that the particles and magnetic field have
the same distribution and that both are reasonably uniform, which may
not be true.
For the very few clusters with such data, the present observations and
this simple argument suggest
that the magnetic field strengths are $\sim 0.5 \, \mu$G, about an order
of magnitude smaller than those derived from Faraday rotation.
This disagreement may indicate that the assumption of a similar and
uniform distribution for the particles and magnetic fields is wrong;
for example, the magnetic field in clusters may be very 
inhomogeneous.
Assuming these uncertain values are correct, the energy density and pressure in
relativistic electrons may be a few percent of values for the thermal
plasma in clusters.
However, the total contribution from relativistic particles is also
uncertain because the ions have not been detected.

\section{Cluster Mergers: the Most Energetic Events Since the Big Bang}
\label{sec:mergers}

There now is considerable evidence that clusters of galaxies and other large
structures in the Universe form hierarchically;
that is, smaller structures form first, and gravity pulls these smaller
structures together to make larger structures.
Clusters of galaxies form by the merger of smaller subclusters and groups
of galaxies.

\begin{figure}
\includegraphics{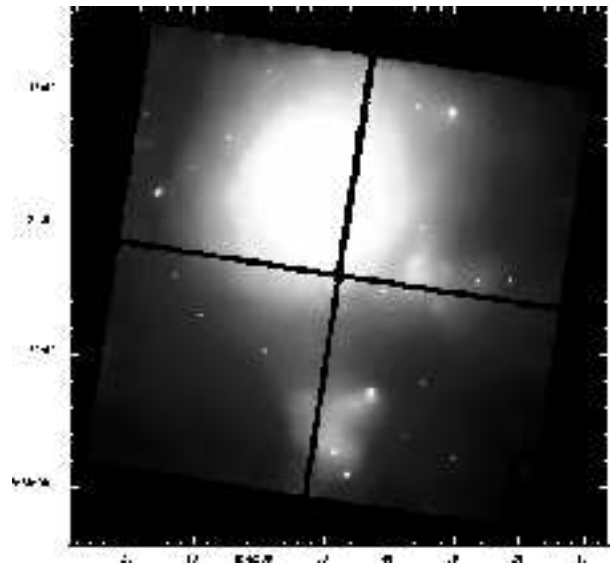}
\caption{
\label{fig:fig3}
The Chandra X-ray image of the merger cluster Abell~85 \cite{ksr02}.
The grey scale burns out the central cooling flow region to show the
outer parts of the cluster.
Two subclusters to the south (lower middle) and southwest (lower right)
are merging with the main cluster.
The southwestern subcluster has a cluster radio relic.
The sharp feature at the northwest of the southern subcluster is a
``cold front''.
}
\end{figure}

Major cluster mergers, in which two subclusters with a total mass
of $\sim 10^{15} \, M_\odot$ collide together at velocities of more
than 2000 km/s, are the most energetic events which have occurred in
the Universe since the Big Bang itself \cite{sar02}.
Cluster mergers release total energies of $\sim 3 \times 10^{64}$ erg.
The motions in cluster mergers are transonic, and the mergers drive shocks
into the intracluster gas.
In major mergers, these merger shocks dissipate total energies of
$\sim 3 \times 10^{63}$ erg.
Such merger shocks are, in fact, the primary heating source of the
intracluster plasma.
For example, Figure~(\ref{fig:fig3}) shows the Chandra image of
the merging cluster Abell~85 \cite{ksr02}.

\begin{figure}
\includegraphics{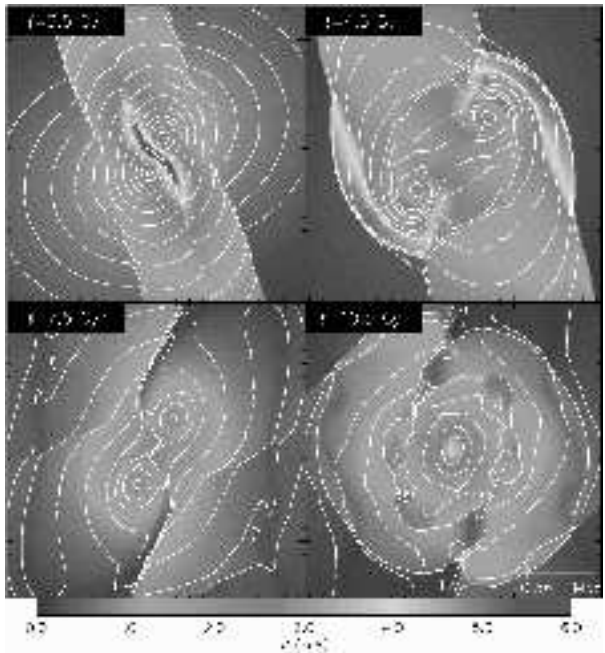}
\caption{
\label{fig:fig4}
The results of a hydrodynamical simulation of a symmetric, off-center
cluster merger \cite{rs01}.
The grey scale shows the temperature, while the contours are the X-ray
surface brightness.
Initially, the shocked region is located between the two subcluster
centers.
Later, the main merger shocks propogate to the outer parts of the cluster,
and other weaker shocks also occur.
By the end of the simulation, the cluster is beginning to return to
equilibrium.
}
\end{figure}

Hydrodynamical simulations of cluster formation and evolution have shown
the importance of merger shocks \cite{rs01}.
The evolution of the structure of merger shocks is illustrated in
Figure~\ref{fig:fig4},
which shows an off-center merger between two symmetric subclusters.
At early stages in the merger (the first panel and earlier), the shocked
region is located between the two subcluster centers and is bounded
on either side by two shocks.
At this time, the subcluster centers, which may contain cooling cores
and central radio sources, are not affected.
Later, these shocks sweep over the subcluster centers (between the first
and second panels).
The main merger shocks pass into the outer parts of the merging system
(panel 2),
and secondary shocks may appear in the inner regions (panel 3).
Eventually, the cluster begins to return to equilibrium (panel 4).

In addition to their thermal effects,
astrophysical shocks at velocities $\ga 1000$ km/s always convert at least
a few percent of the shock energy into the acceleration of relativistic
electrons and ions \cite{be87}.
In general, this occurs through a first-order Fermi acceleration process.
One would thus expect that merger shocks would produce relativistic
electrons, which would be observable through radio synchrotron emission.
In fact, cluster radio relic and cluster radio halo sources are seen
in many clusters (Sec.~\ref{sec:magnetic}).
In every case, these clusters appear to be undergoing a cluster merger;
the cluster Abell~3667 in Figure~(\ref{fig:fig2}) is an example.
Recent Chandra X-ray images indicate that the radio relics lie just
behind merger shocks;
the central radio halos may be due to turbulent particle acceleration after
the passage of the merger shock \cite{mv01}.

\begin{figure}
\includegraphics{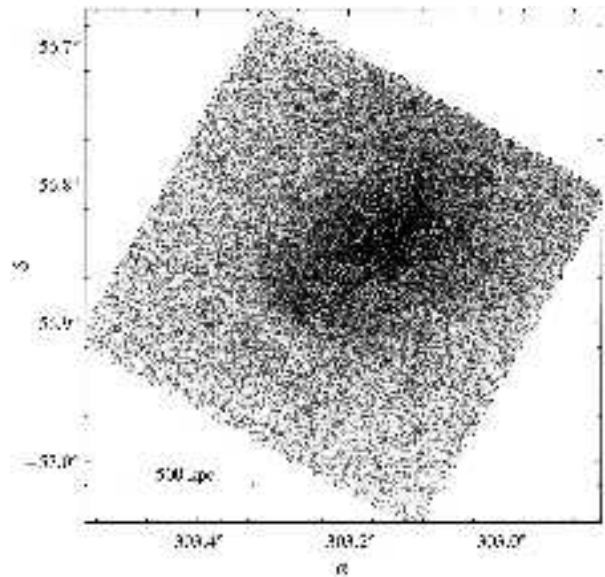}
\caption{
\label{fig:fig5}
A ``cold front'' seen in the Chandra X-ray image of the central region
of the merging cluster Abell~3667 \cite{vmm01}.
[This is the central region of the same cluster shown in
Figure~(\protect\ref{fig:fig2}).]
Note the sharp discontinuity in the X-ray surface brightness.
Spectral analysis show that the brighter gas is actually cooler than the
surrounding fainter gas.
(\copyright 2001, The American Astronomical Society.)
}
\end{figure}

One exciting discovery made with the Chandra X-ray Observatory is the
importance of ``cold fronts'' in merging clusters.
Figure~(\ref{fig:fig5}) shows the cold front seen in the Chandra
X-ray image of the central regions of the merging cluster
Abell~3667 \cite{vmm01};
this is the same cluster shown in Figure~(\protect\ref{fig:fig2}).
The cold front is the sharp surface brightness discontinuity to the
lower right.
When these features were first seen, it was initially assumed that they
merger shocks, and that the brighter inner region resulted from
shock compression.
However, X-ray spectral measurements show that these features are
not shocks.
The temperature of the denser gas is actually lower than that of the
less dense gas in such a way that the pressure is continuous across
the surface brightness discontinuity.
Thus, the specific entropy is actually lower in the denser region.
Shocks are irreversible changes which increase the density, pressure,
and entropy. 
The observed ``cold fronts'' in clusters are not shocks, but rather contact
discontinuities between higher density cool gas and lower density hot gas.

As discussed below (Sec.~\ref{sec:centers}), the centers of many clusters
often contain relative cool ($10^7$ K rather than $10^8$ K) gas.
It is believed that cold fronts occur when clusters containing such
cool cores merger.
The gas in the core is dense enough to survive for some time after
the merger.
As the clusters merge, the cool cores move rapidly through the lower
density shocked gas, producing the cold fronts.

Observations of cluster merger shocks and cold fronts can be used to derive
the kinematics of the merger \cite{msv99,vmm01}.
Most of these diagnostics give the Mach number of the merger
${\cal M}$,
which is the ratio of the merger velocity to the sound speed in
the pre-merger gas.
The Rankine--Hugoniot jump conditions across a merger shock give the
pressure increase across the shock and the shock compression as
\begin{eqnarray}
\frac{P_2}{P_1} & = &
\frac{ 2 \gamma}{\gamma + 1} {\cal M}^2 -
\frac{\gamma - 1}{\gamma + 1} \, \nonumber \\
\frac{v_2}{v_1} =
\frac{\rho_1}{\rho_2} \equiv \frac{1}{C} & = &
\frac{ 2 }{\gamma + 1} \frac{1}{{\cal M}^2} +
\frac{\gamma - 1}{\gamma + 1} \, ,
\label{eq:jumpM}
\end{eqnarray}
where $\gamma = 5/3$ is the adiabatic index for fully ionized plasma,
where $C \equiv \rho_2 / \rho_1$ is the shock compression, and the
subscripts 1 and 2 denote the pre-shock and post-shock gas.
X-ray observations can provide the gas temperature and density on
either side of the shock, and the jump conditions yield ${\cal M}$
and the merger velocity.

For the case of a cold front, the stagnation condition at the leading
edge of the cold front gives \cite{vmm01}
\begin{equation} \label{eq:pst}
\frac{P_{\rm st}}{P_1} = \left\{
\begin{array}{cl}
\left( 1 + \frac{\gamma - 1}{2} {\cal M}^2
\right)^{\frac{\gamma}{\gamma - 1}} \, , &
{\cal M} \le 1 \, , \\
{\cal M}^2 \,
\left( \frac{\gamma + 1}{2}
\right)^{\frac{\gamma + 1}{\gamma - 1}} \,
\left( \gamma - \frac{\gamma - 1}{2 {\cal M}^2} \,
\right)^{- \frac{1}{\gamma - 1}} \, , &
{\cal M} > 1 \, , \\
\end{array}
\right.
\end{equation}
where $P_{\rm st}$ is the pressure at the stagnation point.
If ${\cal M} > 1$ there will be a bow shock ahead of the cold front,
and one can also apply the shock jump conditions,
(Equation~\ref{eq:jumpM}).
The bow shock will be located at some distance (the ``stand off'' distance
$d_s$) ahead of the cold front, and the ratio of this distance to the
radius of curvature of the cold front is a decreasing function of ${\cal M}$.
Finally, the opening angle of the Mach cone formed from the merger shock
will depend on the Mach number as $\theta_M = \csc^{-1} ( {\cal M})$.

Applications of these techniques to observed cluster mergers give
values for the merger Mach number of ${\cal M} \approx 2$ and merger
velocities of $\approx 2000$ km/s \cite{vmm01,ksr02}.

\begin{figure}
\includegraphics{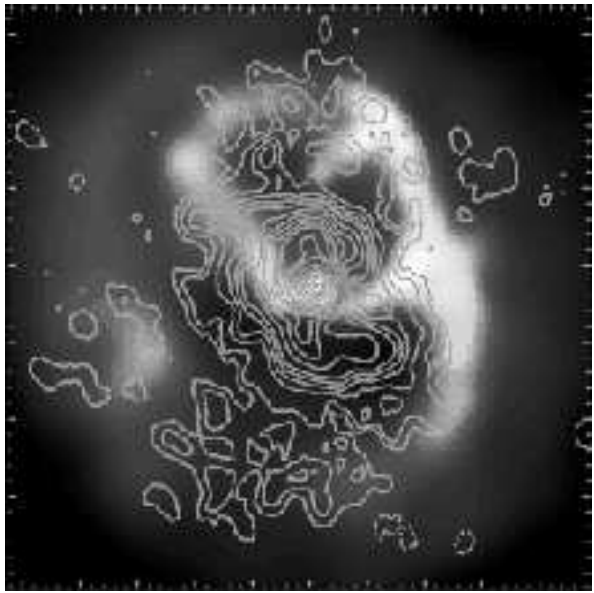}
\caption{
\label{fig:fig6}
The grey scale image is the central $\sim 80\times80$ kpc of the
Chandra X-ray image of the cooling core cluster Abell~2052 \cite{bsm+01}.
The contours are the radio image of the same region.
The hot, X-ray gas is missing from two bubbles to the north (up) and south
(down), and bright X-ray shells surround these bubbles.
The radio image shows that the bubbles are filled by radio emitting plasma
containing relativistic electrons and magnetic fields.
}
\end{figure}

\section{Central Cooling Cores in Clusters}
\label{sec:centers}

At the centers of many clusters of galaxies, the gas temperature is
seen to drop significantly from $\sim 10^8$ K further out to $\sim 10^7$ K
near the center.
At the same time, the gas density rises very rapidly, which gives these
regions very large X-ray surface brightnesses.
The low gas temperatures and high gas densities are the result of
radiative cooling of the gas.
Since the X-ray emissivity varies with the square of the density,
while the thermal energy density is proportional to the gas density,
the X-ray radiation we observe can cool the gas most rapidly in the
dense, central regions of clusters.
In these regions, the radiative cooling times are $\la 3 \times 10^8$ yr,
which is much shorter than the ages of typical clusters of
$\sim 10^{10}$ yr.
As the central gas cools, the weight of the overlying outer gas compresses
the inner gas, resulting in cooler temperatures and very high densities.
I will refer to these cool central regions as ``cooling cores''.

One mystery with previous X-ray observations was whether the gas continues
to cool below $\sim 10^7$ K, and if so, what happens to it.
The rates of gas cooling at higher temperature are quite large,
and corresponding amounts of much cooling gas are not seen.
However, at the center of every cluster with a cooling core, there is a
giant cD galaxy.
These cD galaxies are the largest galaxies seen in the Universe, and are
$\sim$ 10 times larger in mass and radius than other very large galaxies.
The cD galaxies at the centers of cooling flows often contain some cool
gas and some star formation, although in amounts which are $\la 10$\% of
those expected from the rates of radiative cooling of the X-ray gas.

In nearly every case, these central cD galaxies host radio sources.
Recent X-ray observations with the Chandra X-ray Observatory suggest
that interactions between the radio source and the X-ray gas in the
central regions of clusters strongly affects both components.
For example, Figure~(\ref{fig:fig6}) shows the inner region of the
the cluster Abell~2052, which has a cooling core.
The central cD galaxy has the strong radio source 3C~317.
There are central point sources in X-rays and radio which are coincident
with the center of the cD galaxy, which is believed to contain a
supermassive black hole.
The extended radio emission corresponds with ``holes'' in the
X-ray emission, and the radio source is surrounded by a brightened
``shell'' of X-ray emission.
We refer to these structures as ``radio bubbles.''
Similar structures are seen in many other cooling core clusters.

The pressures in the X-ray-bright shells are nearly continuous with the
pressure of the surrounding gas.
There is no clear evidence for strong shocks.
Thus, it seems likely that the radio lobes are displacing and compressing
the X-ray gas, but are, at the same time, confined by the X-ray gas.
The radio bubbles contain relativistic electrons and magnetic
fields, and emit radio synchrotron radiation.

The total energy of the the radio plasma is $\sim 10^{59}$ erg.
If this energy is eventually dissipated into thermal energy in
the X-ray gas, the energy input would be sufficient to balance
cooling for about $10^8$ yr, which is about the radiative cooling time
of the gas.
Thus, energy input from radio sources at the centers of cooling core
clusters may partly balance radiative cooling, and may help to explain
why only a fraction of the X-ray gas cools to low temperatures.

If there were no physical connection between the radio source and
the X-ray emitting plasma in a cluster, then any balance between X-ray
cooling and radio source heating would be a coincidence.
However, one possibility is that the two are coupled, and form a
``feedback loop.''
Observations of nearby examples indicate that all large galaxies,
such as cluster-central cD galaxies, contain supermassive black holes
(SMBHs)
with masses of $\sim 10^8 \, M_\odot$.
Radio sources occur when such a SMBH accretes gas from its environment.
In radio sources, much of the energy release is converted into the
kinetic energy of a pair of oppositely-directed jets of plasma which
expand away from the black hole.
Consider an inactive SMBH black hole at the center of
a cD galaxy in the cooling core of a cluster.
If there is no heat source to balance radiative cooling in the X-ray gas,
it will cool and flow towards the central supermassive black hole.
Part of this gas will be accreted by the supermassive black hole,
and the accretion energy (gravitational binding energy) will power radio
jets.
These will expand into the surrounding X-ray gas, and will inflate two
radio bubbles like those in Figure~(\ref{fig:fig6}).
These bubbles will displace and may eventually heat the X-ray gas, balancing its
radiative cooling.
This will stop the flow of material towards the central SMBH, and
eventually turn off the radio source.
Heating from the radio source will be unable to prevent cooling of
the X-ray gas, and the cycle will start anew.
Various arguments suggest that the radio sources in these systems are
very strongly active for $\sim 10^7$ yr, and that the cycle repeats about
every $\sim 10^8$ yr.

\section{Conclusions}
\label{sec:conclude}

Recent observations and theoretical work indicate that the majority
of the normal, baryonic matter in the low-redshift Universe is in the form
of hot, diffuse plasma.
Clusters of galaxies are particularly striking examples of this.
Large clusters contain $\sim 10^{14} \, M_\odot$ of hot plasma at
a temperature of $\sim 10^8$ K and
typical electron densities of $n_e \sim 10^{-3}$ cm$^{-3}$.
The total thermal energy content of this plasma is $\sim 3 \times 10^{63}$
erg.
This plasma emits most readily in the X-ray band.
Recent X-ray observations with the Chandra X-ray Observatory have confirmed
that the clusters are formed by the merger of smaller structures, and
that merger shocks heat the gas to high temperatures.
Mergers may also accelerate relativistic electrons.
In the central regions of clusters, the hot plasma cools radiatively.
Recent Chandra X-ray Observatory images indicate that the hot thermal
plasma interacts with cluster-central radio sources.

\begin{acknowledgments}
I am grateful to
Huub Rottgering
and
Alexey Vikhlinin
for their very kind permission to use figures from their publications.
Support for this work was provided by the National Aeronautics and Space
Administration through Chandra Award
Numbers 
GO1-2123X,
GO1-2133X,
and
GO2-3159X,
issued by the Chandra X-ray Observatory Center, which is operated by the
Smithsonian Astrophysical Observatory for and on behalf of NASA under
contract NAS8-39073,
and by NASA XMM/Newton Grant NAG5-10075.
\end{acknowledgments}



\begin{thebibliography}{17}
\expandafter\ifx\csname natexlab\endcsname\relax\def\natexlab#1{#1}\fi
\expandafter\ifx\csname bibnamefont\endcsname\relax
  \def\bibnamefont#1{#1}\fi
\expandafter\ifx\csname bibfnamefont\endcsname\relax
  \def\bibfnamefont#1{#1}\fi
\expandafter\ifx\csname citenamefont\endcsname\relax
  \def\citenamefont#1{#1}\fi
\expandafter\ifx\csname url\endcsname\relax
  \def\url#1{\texttt{#1}}\fi
\expandafter\ifx\csname urlprefix\endcsname\relax\def\urlprefix{URL }\fi
\providecommand{\bibinfo}[2]{#2}
\providecommand{\eprint}[2][]{\url{#2}}

\bibitem[{\citenamefont{{Abell} et~al.}(1989)\citenamefont{{Abell}, {Corwin},
  and {Olowin}}}]{aco89}
\bibinfo{author}{\bibfnamefont{G.~O.} \bibnamefont{{Abell}}},
  \bibinfo{author}{\bibfnamefont{H.~G.} \bibnamefont{{Corwin}}},
  \bibnamefont{and} \bibinfo{author}{\bibfnamefont{R.~P.}
  \bibnamefont{{Olowin}}}, \bibinfo{journal}{Astrophys. J. Suppl.}
  \textbf{\bibinfo{volume}{70}}, \bibinfo{pages}{1} (\bibinfo{year}{1989}).

\bibitem[{\citenamefont{{Bahcall}}(2000)}]{nab00}
\bibinfo{author}{\bibfnamefont{N.~A.} \bibnamefont{{Bahcall}}},
  \bibinfo{journal}{Phys. Rep.} \textbf{\bibinfo{volume}{333}},
  \bibinfo{pages}{233} (\bibinfo{year}{2000}).

\bibitem[{\citenamefont{{Sarazin}}(1986)}]{sar86}
\bibinfo{author}{\bibfnamefont{C.~L.} \bibnamefont{{Sarazin}}},
  \bibinfo{journal}{Rev. Mod. Phys.} \textbf{\bibinfo{volume}{58}},
  \bibinfo{pages}{1} (\bibinfo{year}{1986}).

\bibitem[{\citenamefont{{Spitzer}}(1956)}]{spi56}
\bibinfo{author}{\bibfnamefont{L.~J.} \bibnamefont{{Spitzer}}},
  \emph{\bibinfo{title}{{Physics of Fully Ionized Gases}}}
  (\bibinfo{publisher}{Interscience}, \bibinfo{address}{New York},
  \bibinfo{year}{1956}).

\bibitem[{\citenamefont{{Clarke} et~al.}(2001)\citenamefont{{Clarke},
  {Kronberg}, and {B{\" o}hringer}}}]{ckb01}
\bibinfo{author}{\bibfnamefont{T.~E.} \bibnamefont{{Clarke}}},
  \bibinfo{author}{\bibfnamefont{P.~P.} \bibnamefont{{Kronberg}}},
  \bibnamefont{and} \bibinfo{author}{\bibfnamefont{H.}~\bibnamefont{{B{\"
  o}hringer}}}, \bibinfo{journal}{Astrophys. J.}
  \textbf{\bibinfo{volume}{547}}, \bibinfo{pages}{L111} (\bibinfo{year}{2001}).

\bibitem[{\citenamefont{{Giovannini} and {Feretti}}(2002)}]{gf02}
\bibinfo{author}{\bibfnamefont{G.}~\bibnamefont{{Giovannini}}}
  \bibnamefont{and}
  \bibinfo{author}{\bibfnamefont{L.}~\bibnamefont{{Feretti}}}, in
  \emph{\bibinfo{booktitle}{Merging Processes in Galaxy Clusters}}, edited by
  \bibinfo{editor}{\bibfnamefont{L.}~\bibnamefont{{Feretti}}},
  \bibinfo{editor}{\bibfnamefont{I.~M.} \bibnamefont{{Gioia}}},
  \bibnamefont{and}
  \bibinfo{editor}{\bibfnamefont{G.}~\bibnamefont{{Giovannini}}}
  (\bibinfo{publisher}{Kluwer}, \bibinfo{address}{Dordrecht},
  \bibinfo{year}{2002}), pp. \bibinfo{pages}{197--227}.

\bibitem[{\citenamefont{{Rottgering} et~al.}(1997)\citenamefont{{Rottgering},
  {Wieringa}, {Hunstead}, and {Ekers}}}]{rwh+97}
\bibinfo{author}{\bibfnamefont{H.~J.~A.} \bibnamefont{{Rottgering}}},
  \bibinfo{author}{\bibfnamefont{M.~H.} \bibnamefont{{Wieringa}}},
  \bibinfo{author}{\bibfnamefont{R.~W.} \bibnamefont{{Hunstead}}},
  \bibnamefont{and} \bibinfo{author}{\bibfnamefont{R.~D.}
  \bibnamefont{{Ekers}}}, \bibinfo{journal}{Mon.\ Not.\ R. Astron.\ Soc.}
  \textbf{\bibinfo{volume}{290}}, \bibinfo{pages}{577} (\bibinfo{year}{1997}).

\bibitem[{\citenamefont{{Fusco-Femiano}
  et~al.}(1999)\citenamefont{{Fusco-Femiano}, {dal Fiume}, {Feretti},
  {Giovannini}, {Grandi}, {Matt}, {Molendi}, and {Santangelo}}}]{ff+99}
\bibinfo{author}{\bibfnamefont{R.}~\bibnamefont{{Fusco-Femiano}}},
  \bibinfo{author}{\bibfnamefont{D.}~\bibnamefont{{dal Fiume}}},
  \bibinfo{author}{\bibfnamefont{L.}~\bibnamefont{{Feretti}}},
  \bibinfo{author}{\bibfnamefont{G.}~\bibnamefont{{Giovannini}}},
  \bibinfo{author}{\bibfnamefont{P.}~\bibnamefont{{Grandi}}},
  \bibinfo{author}{\bibfnamefont{G.}~\bibnamefont{{Matt}}},
  \bibinfo{author}{\bibfnamefont{S.}~\bibnamefont{{Molendi}}},
  \bibnamefont{and}
  \bibinfo{author}{\bibfnamefont{A.}~\bibnamefont{{Santangelo}}},
  \bibinfo{journal}{Astrophys. J.} \textbf{\bibinfo{volume}{513}},
  \bibinfo{pages}{L21} (\bibinfo{year}{1999}).

\bibitem[{\citenamefont{{Sarazin} and {Lieu}}(1998)}]{sl98}
\bibinfo{author}{\bibfnamefont{C.~L.} \bibnamefont{{Sarazin}}}
  \bibnamefont{and} \bibinfo{author}{\bibfnamefont{R.}~\bibnamefont{{Lieu}}},
  \bibinfo{journal}{Astrophys. J.} \textbf{\bibinfo{volume}{494}},
  \bibinfo{pages}{L177} (\bibinfo{year}{1998}).

\bibitem[{\citenamefont{{Sarazin}}(2002)}]{sar02}
\bibinfo{author}{\bibfnamefont{C.~L.} \bibnamefont{{Sarazin}}}, in
  \emph{\bibinfo{booktitle}{Merging Processes in Galaxy Clusters}}, edited by
  \bibinfo{editor}{\bibfnamefont{L.}~\bibnamefont{{Feretti}}},
  \bibinfo{editor}{\bibfnamefont{I.~M.} \bibnamefont{{Gioia}}},
  \bibnamefont{and}
  \bibinfo{editor}{\bibfnamefont{G.}~\bibnamefont{{Giovannini}}}
  (\bibinfo{publisher}{Kluwer}, \bibinfo{address}{Dordrecht},
  \bibinfo{year}{2002}), pp. \bibinfo{pages}{1--38}.

\bibitem[{\citenamefont{{Kempner} et~al.}(2002)\citenamefont{{Kempner},
  {Sarazin}, and {Ricker}}}]{ksr02}
\bibinfo{author}{\bibfnamefont{J.~C.} \bibnamefont{{Kempner}}},
  \bibinfo{author}{\bibfnamefont{C.~L.} \bibnamefont{{Sarazin}}},
  \bibnamefont{and} \bibinfo{author}{\bibfnamefont{P.~M.}
  \bibnamefont{{Ricker}}}, \bibinfo{journal}{Astrophys. J.}
  \textbf{\bibinfo{volume}{579}}, \bibinfo{pages}{236} (\bibinfo{year}{2002}).

\bibitem[{\citenamefont{{Ricker} and {Sarazin}}(2001)}]{rs01}
\bibinfo{author}{\bibfnamefont{P.~M.} \bibnamefont{{Ricker}}} \bibnamefont{and}
  \bibinfo{author}{\bibfnamefont{C.~L.} \bibnamefont{{Sarazin}}},
  \bibinfo{journal}{Astrophys. J.} \textbf{\bibinfo{volume}{561}},
  \bibinfo{pages}{621} (\bibinfo{year}{2001}).

\bibitem[{\citenamefont{{Blandford} and {Eichler}}(1987)}]{be87}
\bibinfo{author}{\bibfnamefont{R.}~\bibnamefont{{Blandford}}} \bibnamefont{and}
  \bibinfo{author}{\bibfnamefont{D.}~\bibnamefont{{Eichler}}},
  \bibinfo{journal}{Phys. Rep.} \textbf{\bibinfo{volume}{154}},
  \bibinfo{pages}{1} (\bibinfo{year}{1987}).

\bibitem[{\citenamefont{{Markevitch} and {Vikhlinin}}(2001)}]{mv01}
\bibinfo{author}{\bibfnamefont{M.}~\bibnamefont{{Markevitch}}}
  \bibnamefont{and}
  \bibinfo{author}{\bibfnamefont{A.}~\bibnamefont{{Vikhlinin}}},
  \bibinfo{journal}{Astrophys. J.} \textbf{\bibinfo{volume}{563}},
  \bibinfo{pages}{95} (\bibinfo{year}{2001}).

\bibitem[{\citenamefont{{Vikhlinin} et~al.}(2001)\citenamefont{{Vikhlinin},
  {Markevitch}, and {Murray}}}]{vmm01}
\bibinfo{author}{\bibfnamefont{A.}~\bibnamefont{{Vikhlinin}}},
  \bibinfo{author}{\bibfnamefont{M.}~\bibnamefont{{Markevitch}}},
  \bibnamefont{and} \bibinfo{author}{\bibfnamefont{S.~S.}
  \bibnamefont{{Murray}}}, \bibinfo{journal}{Astrophys. J.}
  \textbf{\bibinfo{volume}{551}}, \bibinfo{pages}{160} (\bibinfo{year}{2001}).

\bibitem[{\citenamefont{{Markevitch} et~al.}(1999)\citenamefont{{Markevitch},
  {Sarazin}, and {Vikhlinin}}}]{msv99}
\bibinfo{author}{\bibfnamefont{M.}~\bibnamefont{{Markevitch}}},
  \bibinfo{author}{\bibfnamefont{C.~L.} \bibnamefont{{Sarazin}}},
  \bibnamefont{and}
  \bibinfo{author}{\bibfnamefont{A.}~\bibnamefont{{Vikhlinin}}},
  \bibinfo{journal}{Astrophys. J.} \textbf{\bibinfo{volume}{521}},
  \bibinfo{pages}{526} (\bibinfo{year}{1999}).

\bibitem[{\citenamefont{{Blanton} et~al.}(2001)\citenamefont{{Blanton},
  {Sarazin}, {McNamara}, and {Wise}}}]{bsm+01}
\bibinfo{author}{\bibfnamefont{E.~L.} \bibnamefont{{Blanton}}},
  \bibinfo{author}{\bibfnamefont{C.~L.} \bibnamefont{{Sarazin}}},
  \bibinfo{author}{\bibfnamefont{B.~R.} \bibnamefont{{McNamara}}},
  \bibnamefont{and} \bibinfo{author}{\bibfnamefont{M.~W.}
  \bibnamefont{{Wise}}}, \bibinfo{journal}{Astrophys. J.}
  \textbf{\bibinfo{volume}{558}}, \bibinfo{pages}{L15} (\bibinfo{year}{2001}).

\end{thebibliography}

%
\end{document}